\documentclass{aip-cp}

\usepackage[numbers]{natbib}
\usepackage{rotating}
\usepackage{graphicx}

\newcommand{\ba}{\begin{eqnarray}}
\newcommand{\ea}{\end{eqnarray}}
\newcommand{\ban}{\begin{eqnarray*}}
\newcommand{\ean}{\end{eqnarray*}}
\newcommand{\bsub}{\begin{subequations}}
\newcommand{\esub}{\end{subequations}}

\def\ket#1{|#1\rangle}
\def\bra#1{\langle#1|}
\def\lm{(\lambda,\mu)}
\def\lm0{(\lambda_0,\mu_0)}

\begin{document}

\title{Symmetry Remnants in the Face of\\
Competing Interactions in Nuclei}

\author[aff1]{A. Leviatan\corref{cor1}}
\author[aff2]{M. Macek}
\eaddress{michal.macek@yale.edu}
\affil[aff1]{Racah Institute of Physics, The Hebrew University, 
Jerusalem 91904, Israel}
\affil[aff2]{Center for Theoretical Physics, Sloane Physics Laboratory, 
Yale University, New Haven, CT 06520-8120, USA}
\corresp[cor1]{Corresponding author: ami@phys.huji.ac.il}

\maketitle

\begin{abstract}
Detailed description of nuclei necessitates model Hamiltonians 
which break most dynamical symmetries. Nevertheless, 
generalized notions of partial and quasi dynamical symmetries may still be 
applicable to selected subsets of states, amidst a complicated environment 
of other states. We examine such scenarios in 
the context of nuclear shape-phase transitions.
\end{abstract}

\section{INTRODUCTION}
The inherent simplicity encoded in the pattern of collective states in nuclei, 
highlights the role of underlying symmetries in explaining their structure. 
Notable examples are the U(5), SU(3) and O(6) dynamical symmetries of 
the interacting boson model (IBM)~\cite{ibm}, which act as benchmarks 
for spherical-vibrator, axial-rotor and $\gamma$-unstable types of 
collective motion. Each dynamical symmetry (DS) provides a physically 
transparent analytic solution and good quantum numbers for all states of 
the system. The majority of nuclei, 
however, exhibit strong deviations from these solvable limits.
Often, in spite of the required symmetry-breaking terms 
in the Hamiltonian, some eigenstates appear to obey the symmetry 
while other states do not. 
To accommodate such more realistic scenarios, one needs to 
enlarge the traditional concept of exact symmetries. In the present 
contribution, we discuss two such generalizations, partial and 
quasi dynamical symmetries, and show their ability to characterize 
the remaining regularity amidst a complicated 
(at times chaotic) environment 
of other states, a situation 
encountered in a nuclear shape-phase transition.

\section{PARTIAL AND QUASI DYNAMICAL SYMMETRIES}
Partial dynamical symmetry (PDS) corresponds to a situation in which 
only part of the eigenspectrum 
is solvable and/or has good symmetry-based quantum 
numbers~\cite{lev96,lev11}.
Algorithms for constructing Hamiltonians 
with PDS are available~\cite{lev11} 
and empirical evidence for their occurrence 
in a wide range of nuclei, has been presented~[2-8]. 

PDS reflects the purity of selected eigenstates with respect to a DS basis.
Its quantitative measure is based on the concept of Shannon entropy.
Consider an eigenfunction of the IBM Hamiltonian, $\ket{L}$, with 
angular momentum $L$, and denote by $C^{(L)}_{n_d,\tau,n_{\Delta}}$ 
and $C^{(L)}_{(\lambda,\mu),K}$ its expansion coefficients in the 
U(5) and SU(3) bases, respectively. 
Here $n_d,\tau,(\lambda,\mu)$ denote U(5), O(5) and SU(3) 
irreducible representations (irreps) 
and ($n_{\Delta},K$) are multiplicity labels. 
The U(5) ($n_d$) and SU(3) $[(\lambda,\mu)]$ 
probability distributions, $P_{n_d}^{(L)}$ and $P_{(\lambda,\mu)}^{(L)}$, 
and Shannon entropies, $S_{\rm U5}(L)$ and $S_{\rm SU3}(L)$, are given~by
\begin{equation}
P_{n_d}^{(L)} = \sum_{\tau,n_{\Delta}}\vert C^{(L)}_{n_d,\tau,n_{\Delta}}\vert^2
\;\;, \;\;
P_{(\lambda,\mu)}^{(L)} = \sum_{K}
\vert C^{(L)}_{(\lambda,\mu),K}\vert^2 \;\; , \;\;
S_\mathrm{G}(L) = 
-\frac{1}{\ln D_G}
\sum_{\alpha} P_{\alpha}^{(L)} \ln P_{\alpha}^{(L)} ~,
\label{Prob}
\end{equation}
where $[G={\rm U(5),\,SU(3)}]$ and $D_G$ counts the number of 
possible $G$-irreps for a given boson-number $N$ and $L$. 
$S_{G}(L)$ vanishes when the considered state 
is pure with good $G$-symmetry
[$S_\mathrm{G}(L)\!=\!0$], and is positive for a mixed state. 
The maximal value [$S_\mathrm{G}(L)\!=\!1$] 
is obtained when $\ket{L}$ is uniformly spread among the irreps of $G$, 
{\it i.e.}, for $P_{G}^{(L)}\!=\!1/D_G$. 
Intermediate values, $0 < S_\mathrm{G}(L) < 1$, 
indicate partial fragmentation 
of the state $\ket{L}$ in the respective DS basis.

Quasi dynamical symmetry (QDS) corresponds to the situation for 
which selected eigenstates of a Hamiltonian with broken symmetries, 
continue to exhibit characteristic properties 
({\it e.g.}, energy and B(E2) ratios) of the closet 
DS limit~\cite{rowe04}. 
This ``apparent'' symmetry is due to a coherent mixing of irreps, 
imprinting an adiabatic motion~\cite{MDSC10}.

In case of SU(3)-QDS, the coherent mixing arises 
when the SU(3) expansion coefficients, $C^{(L)}_{(\lambda,\mu),K}$, 
are approximately independent of $L$ for a class of states. 
Such strong correlations between different $L$ states,  
signal a common intrinsic structure and the formation of a 
rotational band. 
Focusing, for example, on the $L\!=\!0,2,4,6$, members of $K\!=\!0$ bands,
given a $L=0^{+}_i$ state, among the ensemble of possible states, 
we associate with it those $L_j>0$ states which show the maximum 
correlation, $\max_{j}\{\pi(0_i,L_j)\}$. 
A unit value, $\pi(0_i,L_j)=1$, for the Pearson coefficient
implies a perfect correlation. 
To quantify the amount of coherence (hence of SU(3)-QDS) in the chosen 
set of states, we consider the product 
$C_{\rm SU3}(0_i{\rm -}6) \equiv 
\max_{j}\{\pi(0_i,2_j)\}\,
\max_{k}\{\pi(0_i,4_k)\}\,
\max_{\ell}\{\pi(0_i,6_{\ell})\}$~\cite{Macek10}. 
The set of states $\{0_i,\,2_j,\,4_k,\,6_{\ell}\}$ is considered as 
comprising a $K=0$ band with SU(3)-QDS,
if $C_{\rm SU3}(0_i{\rm -}6)\approx 1$.

\section{ORDER AMIDST CHAOS AND PERSISTING SYMMETRIES} 
The presence of terms in the Hamiltonian with incompatible 
(non-commuting) symmetries, has a profound effect on the 
possible emergence of quantum phase transitions (QPT). 
The latter are structural changes in the properties 
of the system induced by a variation of parameters in the Hamiltonian. 
Such ground state transformations are manifested empirically in nuclei 
as transitions between different shapes~\cite{cejnar10}. 
The competing interactions 
that drive the QPT can affect dramatically the nature of the dynamics 
and, in some cases, lead to 
an intricate interplay of order and chaos~[13-15]. 
Even in such circumstances, generalized symmetries such as PDS and QDS 
can endure, and be assigned to particular regular multiplets of states 
which survive amidst a complicated environment of other 
states~\cite{maclev14}.
\begin{figure}[t]
  \centerline{
\includegraphics[width=0.95\linewidth]{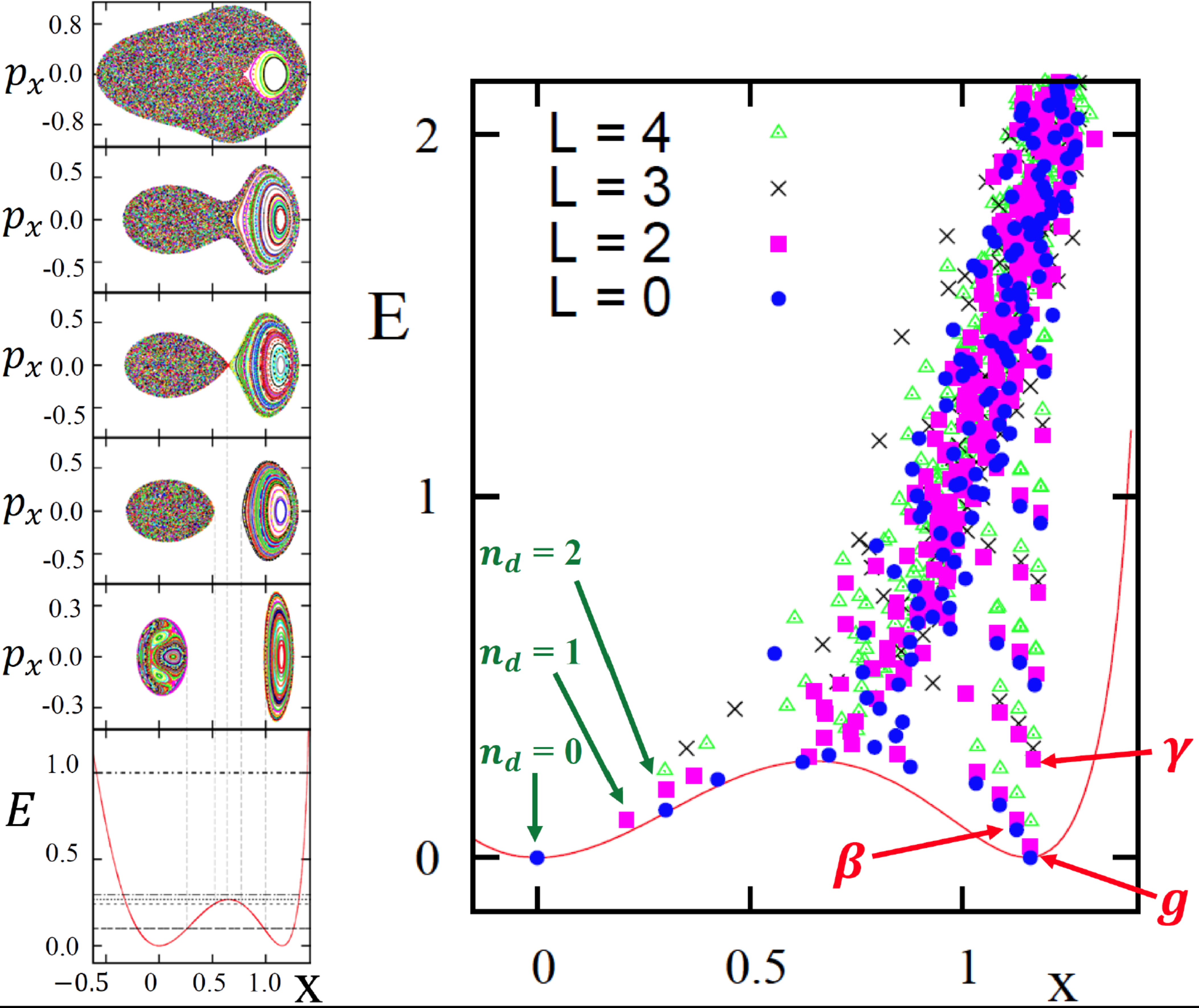}}
  \caption{
Left panel: Poincar\'e sections for the classical critical-point dynamics 
(upper five rows) evaluated at five consecutive energies (indicated by 
horizontal lines in the bottom row on top of the classical potential). 
Right panel: Peres lattices $\{x_i,E_i\}$ for $L=0,2,3,4$ and $N=50$ 
eigenstates of $\hat{H}_{\rm cri}$, Eq.~(\ref{Hcri}), 
overlayed on the classical potential. 
Regular $n_d$-multiplets of spherical type of states and a few 
regular $K$-bands of deformed type of states, are marked by arrows.
Adapted from~\cite{maclev14}.}
\end{figure}
\begin{figure}[t]
  \centerline{
\includegraphics[width=\linewidth]{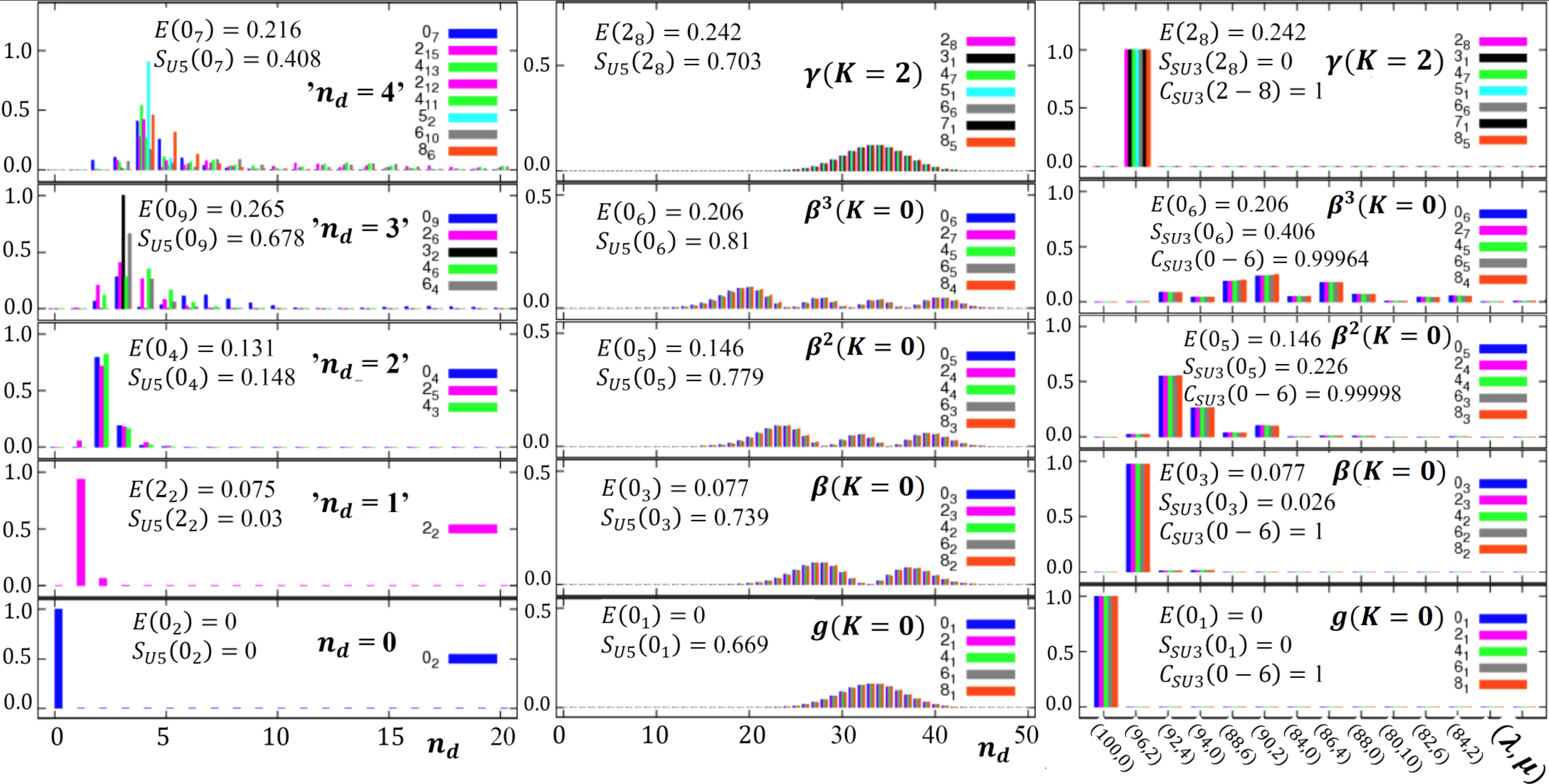}}
\caption{U(5) $(n_d)$ and SU(3) $[(\lambda,\mu)]$ 
decomposition for the regular states shown in Fig.~1. 
Left column: $P_{n_d}^{(L_i)}$ for spherical type of states arranged 
in U(5)-like $n_d$-multiplets. 
Middle column: $P_{n_d}^{(L_i)}$ for deformed type of states arranged in 
rotational $K$-bands. 
Right column: $P_{(\lambda,\mu)}^{(L_i)}$ for the same deformed type of states. 
Shannon entropies $S_{\rm U5}(L_i)\approx 0$ and $S_{\rm SU3}(L_i)\approx 0$ 
signal U(5)-PDS and SU(3)-PDS, respectively. 
SU(3) Pearson correlator $C_{\rm SU3}(0{\rm -}6)\approx 1$, 
signals SU(3)-QDS. Adapted from~\cite{maclev14}.}
\end{figure}

Focusing on the intrinsic dynamics~\cite{lev87} 
at the critical-point of a first-order 
QPT between spherical [U(5)] and deformed [SU(3)] shapes, the relevant 
IBM Hamiltonian can be transcribed in the 
form~\cite{Lev06,Lev07}
\begin{equation}
\hat{H}_{\rm cri}= 
h_{2}\,P^{\dagger}_{2}\cdot \tilde{P}_{2}\;\;\;,\;\;\;
P^{\dagger}_{2m} = 2d^{\dagger}_{m}s^{\dagger} + 
\sqrt{7}\, (d^{\dagger}\,d^{\dagger})^{(2)}_{m} \;\;\; , \;\;\;
\tilde P_{2m} = (-)^{m}P_{2,-m} ~,
\label{Hcri}
\end{equation}
where $s^{\dag}$ ($d^{\dag}_{m}$) are monopole (quadrupole) bosons.
$\hat{H}_{\rm cri}$ mixes terms from different DS chains of the IBM, 
hence is non-integrable. The corresponding classical Hamiltonian, 
obtained by Glauber coherent states, has a Landau potential with two 
degenerate spherical and prolate-deformed minima, as shown in Fig.~1.

For $L=0$, the classical Hamiltonian is two-dimensional and its 
dynamics can be depicted conveniently via 
Poincar\'e surfaces of sections. 
These $(p_x,x)$ sections in the plane $y=0$, are shown in 
the left column of Fig.~1 for representative energies. 
The dynamics in the region of the deformed minimum ($x\approx 1$) 
is seen to be robustly regular. The trajectories form a single island 
and remain regular even 
at energies far exceeding the barrier height. 
In contrast, the dynamics in the region of the spherical minimum 
($x\approx 0$) shows a change with energy 
from regularity to chaos, until complete chaoticity is reached 
near the barrier top. 
The clear separation between regular and chaotic dynamics, 
associated with the two minima, persists all the way to 
the barrier energy, where the two regions just touch. 
At higher energy, 
a layer of chaos develops in the deformed region and gradually dominates 
the surviving regular island.

A quantum analysis is based on Peres lattices~\cite{Peres84}.
The latter are constructed by plotting the expectation 
values $x_i \equiv \sqrt{2 \bra{i}\hat{n}_d\ket{i}/N}$
of the $d$-boson number operator, 
versus the energy $E_i = \bra{i}\hat{H}\ket{i}$ 
of the Hamiltonian eigenstates $\ket{i}$. 
The lattices $\{x_i,E_i\}$ corresponding to 
regular dynamics display an ordered pattern, while chaotic 
dynamics leads to disordered meshes of points~\cite{Peres84,Stran09}. 
The quantity $x_i$ is related to the coordinate $x$ in the classical 
potential, hence the indicated lattices can distinguish regular from 
irregular states and associate them with a given region in phase space.

The Peres lattices  
corresponding to 
($N\!=\!50,\,L\!=\!0,2,3,4$) eigenstates of $\hat{H}_\mathrm{cri}$
are shown in the right column of Fig.~1, 
overlayed on the classical  potential, $V(x,y=0)$. 
They disclose 
regular sequences of states localized within and above the deformed well. 
They are comprised of rotational states with $L=0,2,4,\ldots$ 
forming regular $K\!=\!0$ bands and 
sequences $L=2,3,4,\ldots$ 
forming $K=2$ bands. 
Additional $K$-bands (not shown in Fig.~1), 
corresponding to multiple $\beta$ and $\gamma$ vibrations 
about the deformed shape, can also be identified. 
Such ordered band-structures persist to energies above the barrier and 
are not present in the disordered (chaotic) portions 
of the Peres lattice. 
At low-energy, in the vicinity of the spherical well, one can also detect 
multiplets of states with $L=0$, $L=2$ and $L=0,2,4$, typical of 
quadrupole excitations with $n_d=0,1,2,$ of a spherical shape.

The nature of the surviving regular sequences of selected states 
is revealed in a symmetry analysis of their wave functions. 
The left column of Fig.~2 shows the U(5) $n_d$-probabilities 
$P_{n_d}^{(L_i)}$, Eq.~(\ref{Prob}), for eigenstates of $\hat{H}_{\rm cri}$, 
selected on the  basis of having 
the largest components with $n_d=0,1,2,3,4$,
within the given $L$ spectra. 
The states are arranged into panels labeled 
by `$n_d$' to conform with the structure of the $n_d$-multiplets of the 
U(5) DS limit. The U(5) Shannon entropy $S_{\rm U5}(L_i)$, Eq.~(\ref{Prob}), 
is indicated for representative eigenstates. 
In particular, the zero-energy $L\!=\!0^{+}_2$ state 
is seen to be a pure $n_d\!=\!0$ state, 
with $S_{\rm U5}\!=\!0$. 
It is a solvable eigenstate of $\hat{H}_{\rm cri}$, 
exemplifying U(5)-PDS~\cite{Lev07}. 
The state $2^{+}_2$ has a pronounced $n_d\!=\!1$ 
component~(96\%) and the states ($L=0^{+}_4,\,2^{+}_5,\,4^{+}_3$) 
in the third panel, have a pronounced $n_d\!=\!2$ component and
a low value of $S_{\rm U5}< 0.15$.
All the above states with $`n_d\leq 2$' have a dominant single $n_d$ 
component, and hence qualify as `spherical' type of states. 
They are marked by their $n_d$-values in the Peres lattices of Fig.~1. 
In contrast, the states in the panels `$n_d=3$' and `$n_d=4$' of Fig.~2, 
are significantly fragmented. A notable exception is 
the $L=3^{+}_2$ state, which is a solvable U(5)-PDS eigenstate 
with $n_d=3$~\cite{Lev07}.
The existence in the spectrum of specific spherical-type of states with 
either $P_{n_d}^{(L)}\!=\!1$ $[S_{\rm U5}(L)\!=\!0]$ 
or $P_{n_d}^{(L)}\approx 1$ $[S_{\rm U5}(L)\approx 0]$, exemplifies 
the presence of an exact or approximate U(5) PDS at the critical-point.

The states considered in the middle and right columns of Fig.~2 
have a different character. They 
belong to the five lowest regular sequences 
seen in the Peres lattices of Fig.~1, in the region $x\geq 1$. 
As shown in the middle column of Fig.~2,
they have a broad $n_d$-distribution $P_{n_d}^{(L_i)}$ 
and large $S_{\rm U5}(L)> 0$, 
hence are qualified as `deformed'-type of states, 
forming rotational bands:
$g(K\!=\!0),\,\beta(K\!=\!0),\,\beta^2(K\!=\!0),
\,\beta^3(K\!=\!0)$ and $\gamma(K\!=\!2)$.
Each panel in the right column of Fig.~2, 
depicts the SU(3) $(\lambda,\mu)$-distribution $P_{(\lambda,\mu)}^{(L_i)}$ 
for the band members, the 
SU(3) Shannon entropy $S_{\rm SU3}(L)$, Eq.~(\ref{Prob}), 
for the bandhead state, and the Pearson correlator 
$C_{\rm SU3}(0_i{\rm -}6)$, mentioned above. 
The ground $g(K\!=\!0)$ and $\gamma(K\!=\!2)$  
bands are pure $[S_{\rm SU3}=0$] 
with $(\lambda,\mu) = (2N,0)$ and $(2N-4,2)$ SU(3) character, 
respectively. These are solvable bands of $\hat{H}_{\rm cri}$ and 
exemplify SU(3)-PDS~\cite{Lev07}. 
The non-solvable $K$-bands are mixed with respect to SU(3) 
in a coherent, $L$-independent manner, hence exemplify SU(3)-QDS. 
As expected, $C_{\rm SU3}(0_i{\rm -}6)\approx 1$ for these regular 
$K$-bands. 

The persistence of regular U(5)-like and SU(3)-like multiplets
typifies ``emergent simplicity out of~complexity'' and 
demonstrates the potential relevance of PDS and QDS in characterizing 
symmetry remnants in the face of competing interactions in nuclei. 
This work is supported by the Israel Science Foundation.

\nocite{*}
\bibliographystyle{aipnum-cp}%

\end{document}